\documentclass[12pt]{iopart}

\usepackage{graphicx}
\begin{document}

\title{Neutrinos: Theory and Phenomenology}

\author{Stephen Parke\footnote{Presentation at the ``Nobel Symposium on LHC results'', May 13-19, 2013 at Krusenberg, Uppsala, Sweden.}}

\address{Theoretical Physics Dept., Fermi National Accelerator Laboratory,\\ P.~O.~Box 500, Batavia, IL 60510,USA.}\ead{parke@fnal.gov}
\begin{abstract}
The theory and phenomenology of neutrinos will be addressed, especially that relating to the observation of neutrino flavor transformations. The current status and implications for future experiments will be discussed with special emphasis on the experiments that will determine the neutrino mass ordering, the dominant flavor content of the neutrino mass eigenstate with the smallest electron neutrino content and the size of CP violation in the neutrino sector. Beyond the neutrino Standard Model, the evidence for and a possible definitive experiment to confirm or refute the existence of light sterile neutrinos will be briefly discussed.

\end{abstract}

\maketitle

\section{Introduction}

Fifteen years ago this year, the SuperKamiokande collaboration presented a talk titled ``Evidence for $\nu_\mu$ Oscillations'' at the  Neutrino 1998 conference \cite{Kajita:1998nk}. This set the particle physics world ``abuzz'' since, if neutrinos change flavor,  it implies that their clocks are ticking and therefore they cannot be traveling at the speed of light. Hence neutrinos have mass.

Fast forward fifteen years and the evidence for neutrino flavor conversion is overwhelming.  The simplest and only satisfactory description of all the data is that neutrinos have distinct masses and mix.
Two distinct baseline (L) divided by neutrino energy (E) scales have been identified corresponding to two distinct $\delta m^2_{ij}\equiv m^2_i-m^2_j$  for the neutrino mass eigenstates\footnote{ The LSND, miniBooNE, reactor and source anomalies, which do not fit this paradigm, will be addressed in a later section}: 
\begin{eqnarray}
&L/E = 500  ~{\rm km/GeV}  \quad &{\rm and}  \quad \delta m_{atm}^2 =2.4 \times 10^{-3}  ~{\rm eV}^2 \nonumber \\  
&L/E = 15,000 ~{\rm km/GeV}   \quad& {\rm and}  \quad \delta m_{sol}^2 =7.5 \times 10^{-5} ~{\rm eV}^2. \nonumber
\end{eqnarray} 
These are known as the atmospheric and solar scales, respectively. 

Since it is most likely that the Higgs boson has been discovered at the LHC, it is natural to ask how the neutrinos couple to the Higgs boson. First, what is ``mass'' for a fermion? It is a coupling of the right and left components of the field, and for the neutrino this coupling depends on whether the neutrino is a Dirac particle, like all the other fermions in the Standard Model or a Majorana fermion, which would make the neutrino unique amongst the particles of the Standard Model. The couplings for both Dirac and Majorana  \cite{Weinberg:1979sa} neutrinos are given in the following Table:\\

\hspace*{-0.8cm}\begin{tabular}{rcccl}
{\bf Type:}   &  {\bf Mass Term}	 & {\bf Coupling to Higgs} &{\bf  \# comp. } & {\bf Lepton Number} \\[2mm] 
{\bf  Dirac:}   &  $\bar{\nu}_R \nu_L + \bar{\nu}_L \nu_R$ & $\bar{L}H\nu_R$ &  4 & Conserved \\[2mm]
{\bf Majorana:} &  $\overline{\nu_L}\nu^c_L$ & $\frac{1} {M}(\bar{L}H)^2$ &  2  &  Violated\\[5mm]
\end{tabular}

Determining whether the nature of the neutrino is Dirac or Majorana is one of the big unanswered questions in neutrino physics and is being addressed through neutrinolesss, double beta decay experiments.
Independent of the nature of the neutrino, the partial width of the Higgs decaying to two massive neutrinos is given by
\begin{eqnarray}
\Gamma_{\rm tree}(H\rightarrow \nu_i \bar{\nu}_i) \approx \left( \frac{m_{\nu_i}}{m_\tau} \right)^2 \Gamma(H\rightarrow \tau \bar{\tau})  \approx 10^{-20} ~\Gamma(H\rightarrow \tau \bar{\tau})
\end{eqnarray}
So not only is this decay invisible, it is impossibly tiny and swamped by other invisible decays of the Higgs, such as  $H \rightarrow ZZ \rightarrow 4 \nu$!

In seesaw models, where the mass of the neutrinos is naturally very light, it is possible that LHC could see physics beyond the SM, such as right handed heavy neutrinos or doubly charged scalar Higgs particles, if the mass scale is within reach of this collider.

\section{Neutrino Masses and Mixings}
The three known neutrino flavor states, $\nu_{e}, \nu_\mu, \nu_\tau$, and the three neutrino mass eigenstates, $\nu_{1}, \nu_2, \nu_3$, are
related as follows:
\begin{eqnarray}
\left( \begin{array}{l}
 \nu_{e} \\ \nu_\mu \\ \nu_\tau \end{array}  \right) =
 \left( \begin{array}{lll}
 U_{e 1} & U_{e 2} & U_{e 3} \\
  U_{\mu 1} & U_{\mu 2} & U_{\mu 3} \\
   U_{\tau 1} & U_{\tau 2} & U_{\tau 3} 
   \end{array}  \right)
   \left( \begin{array}{l}
 \nu_{1} \\ \nu_2 \\ \nu_3 \end{array}  \right) 
 \end{eqnarray}
 where the U matrix is unitary and referred to as the PMNS matrix.  The mass eigenstates are labelled such that 
 $%\begin{eqnarray}
 |U_{e1}|^2 > |U_{e2}|^2 > |U_{e3}|^2, 
 $%\end{eqnarray}
  which implies that, by definition, the
 \begin{center}
 $\nu_e$ component of $\nu_1  \quad > \quad  \nu_e$ component  of $\nu_2  \quad $  $ > \quad \nu_e$ component of 
 $\nu_3 $.
  \end{center}

\subsection{Masses} 
   \begin{figure*}[b]
\begin{center}
 \includegraphics[angle=0, scale= .35]{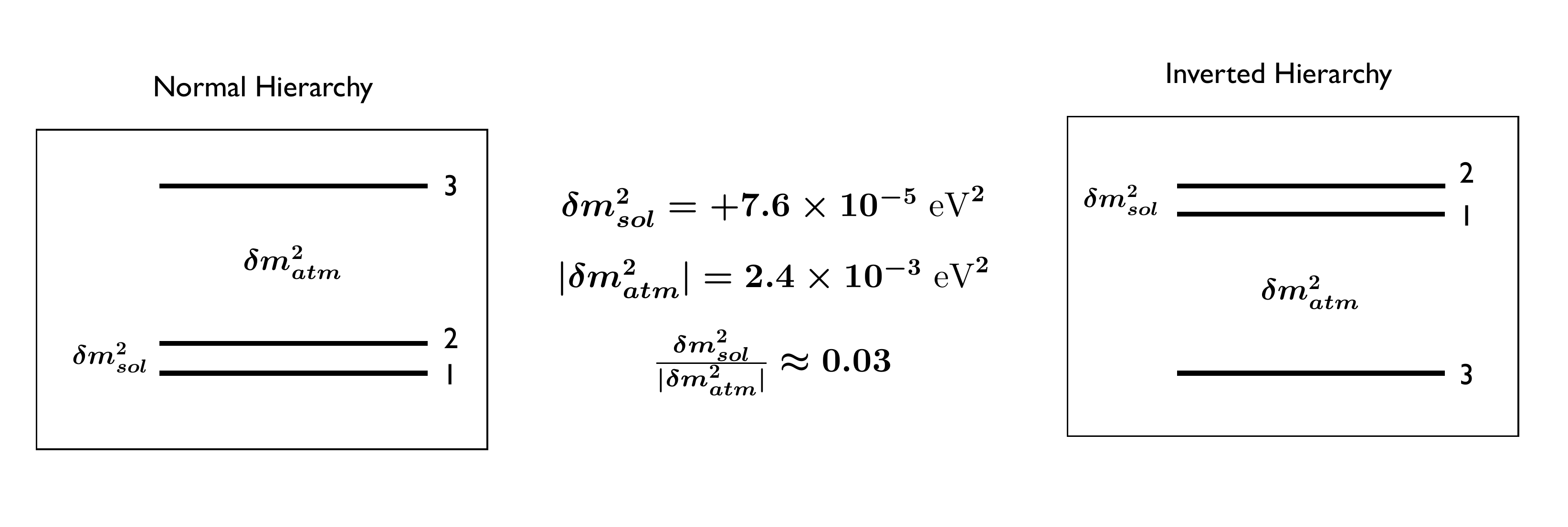}
\end{center}
\caption{What is known about the square of the neutrino masses for the two atmospheric mass hierarchies.
}
\label{fig:mass0}
\end{figure*}

%  \begin{figure*}[b]
%\begin{center}
% \includegraphics[angle=0, scale= .43]{mass.pdf}
%\end{center}
%\caption{What is known about the masses of the neutrinos.
%}
%\label{fig:mass}
%\end{figure*}

 With this choice of labeling of the neutrino mass eigenstates, the solar oscillations are governed by $\delta m^2_{21}$ as both $\nu_1$ and $\nu_2$ have a significant $\nu_e$ component. 
 Whereas the atmospheric oscillations are governed by 
$\delta m^2_{31} \approx \delta m^2_{32}$ as $\nu_3$ has a small $\nu_e$ component required by the small $\nu_e$ involvement shown by the results of the SuperKamiokande and Chooz experiments.  The mass ordering of $\nu_1$ and $\nu_2$ was determined by matter effects in the interior of the sun by the SNO experiment \cite{Ahmed:2003kj}. Their measurement of the charge current to neutral current ratio of less than one half, for the $^8$B high energy solar neutrinos, implies that the higher mass state has the lower $\nu_e$ component i.e. $m^2_2 > m^2_1$  or $\delta m^2_{21}>0$. 

\begin{figure*}[t]
\begin{center}
 \includegraphics[angle=0, scale= .40]{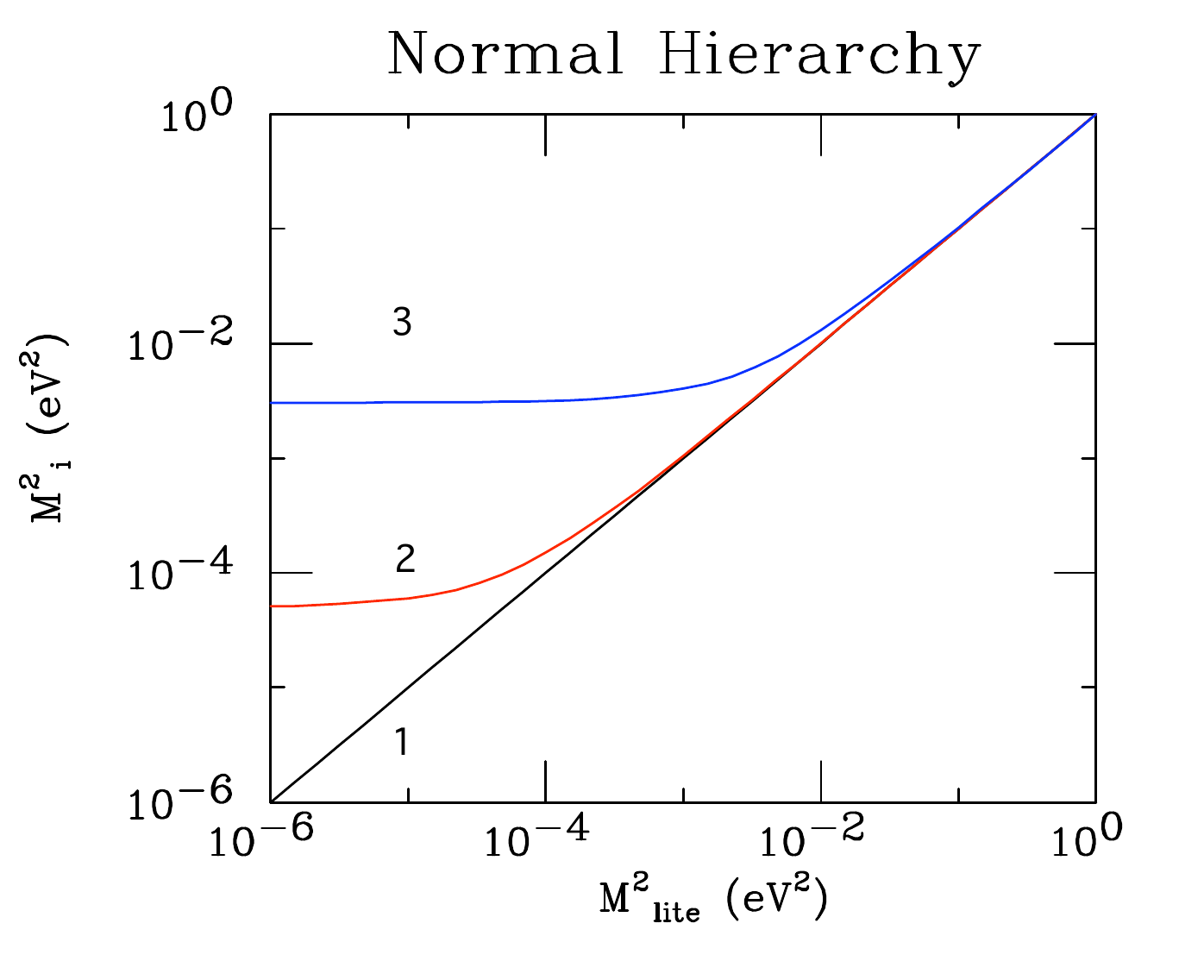}
 \includegraphics[angle=0, scale= .40]{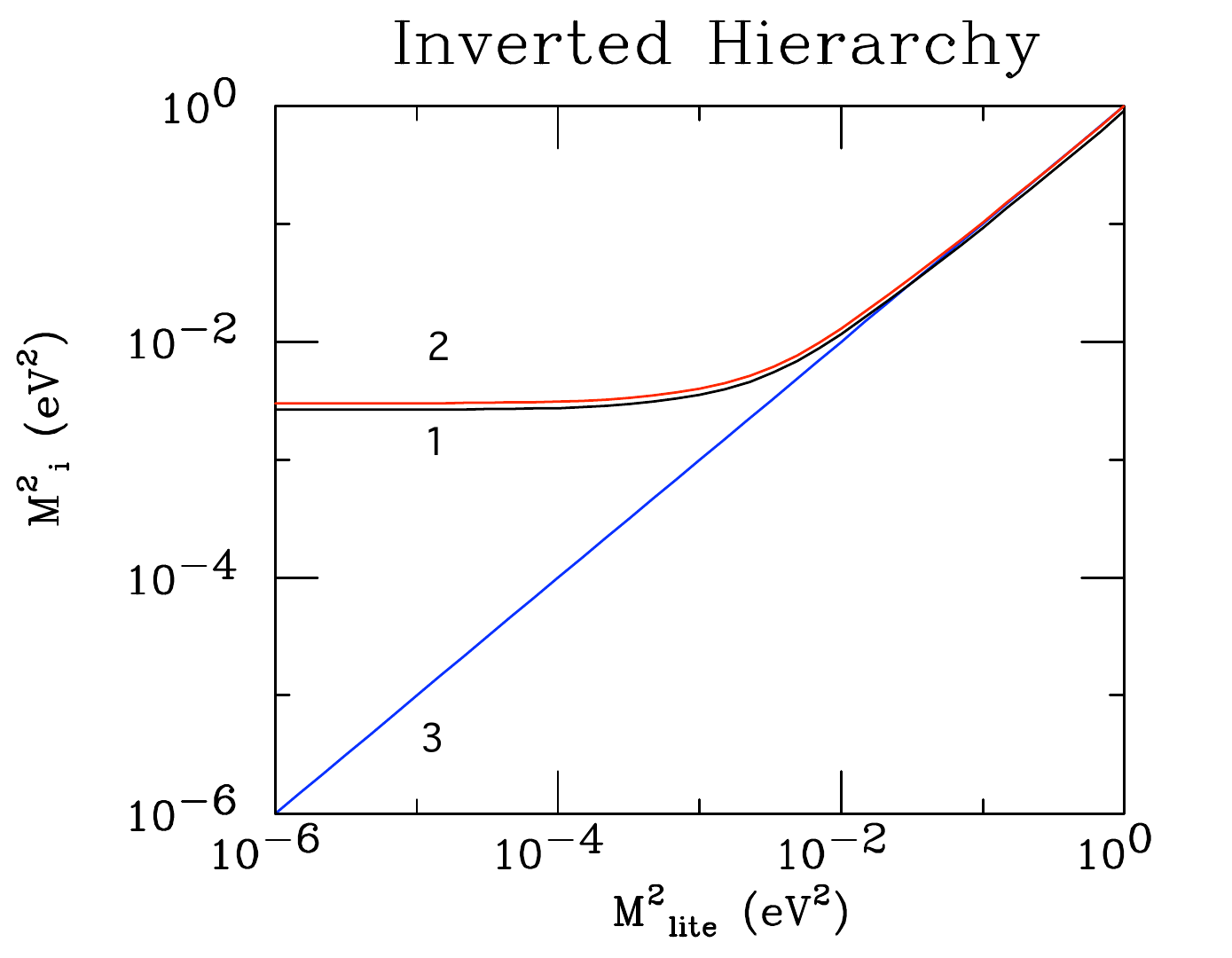}
 \end{center}
\caption{Left panel is the neutrino masses squared as a function of the lightest neutrino mass for the normal hierarchy, here 
$m^2_{\rm lite}= m^2_1$. Right panel is for the inverted hierarchy, where $m^2_{\rm lite}= m^2_3$.
The neutrino mass eigenstate $\nu_1$ and $\nu_2$ are electron neutrino rich whereas $\nu_3$ has only a small electron neutrino component.
}
\label{fig:mass}
\end{figure*}

The atmospheric neutrino mass ordering, 
%\begin{eqnarray}
$m^2_3 > {\rm or} < m^2_2, m^2_1$ % \nonumber
%\end{eqnarray}
 is still to be determined, see Fig. \ref{fig:mass0}.  If $m^2_3 > m^2_2$, the ordering is known as the normal hierarchy (NH), whereas if  $m^2_3 < m^2_1$ the ordering is known as the inverted hierarchy (IH). Fig. \ref{fig:mass} shows the masses as a function of the lightest neutrino mass.

 The sum of the masses of the neutrinos satisfies
 \begin{eqnarray}
 \sqrt{\delta m^2_{atm}} =0.05 ~{\rm eV} < \sum m_{\nu_i} < 0.5 ~{\rm eV}.
 \end{eqnarray}
So the $\sum m_{\nu_i} $ ranges from $10^{-7} $ to  $10^{-6} $ times $m_e$, however the mass of the lightest neutrino, $m_{lite}$, could be very small. If $m_{lite} \ll \sqrt{\delta m^2_{sol}} \sim 0.01~{\rm eV}^2$, then this is an additional scale to be explained by a theory of neutrino masses and mixings. 

\subsection{Mixings}
\begin{figure*}[ht]
\begin{center}
 \includegraphics[angle=0, scale= .53]{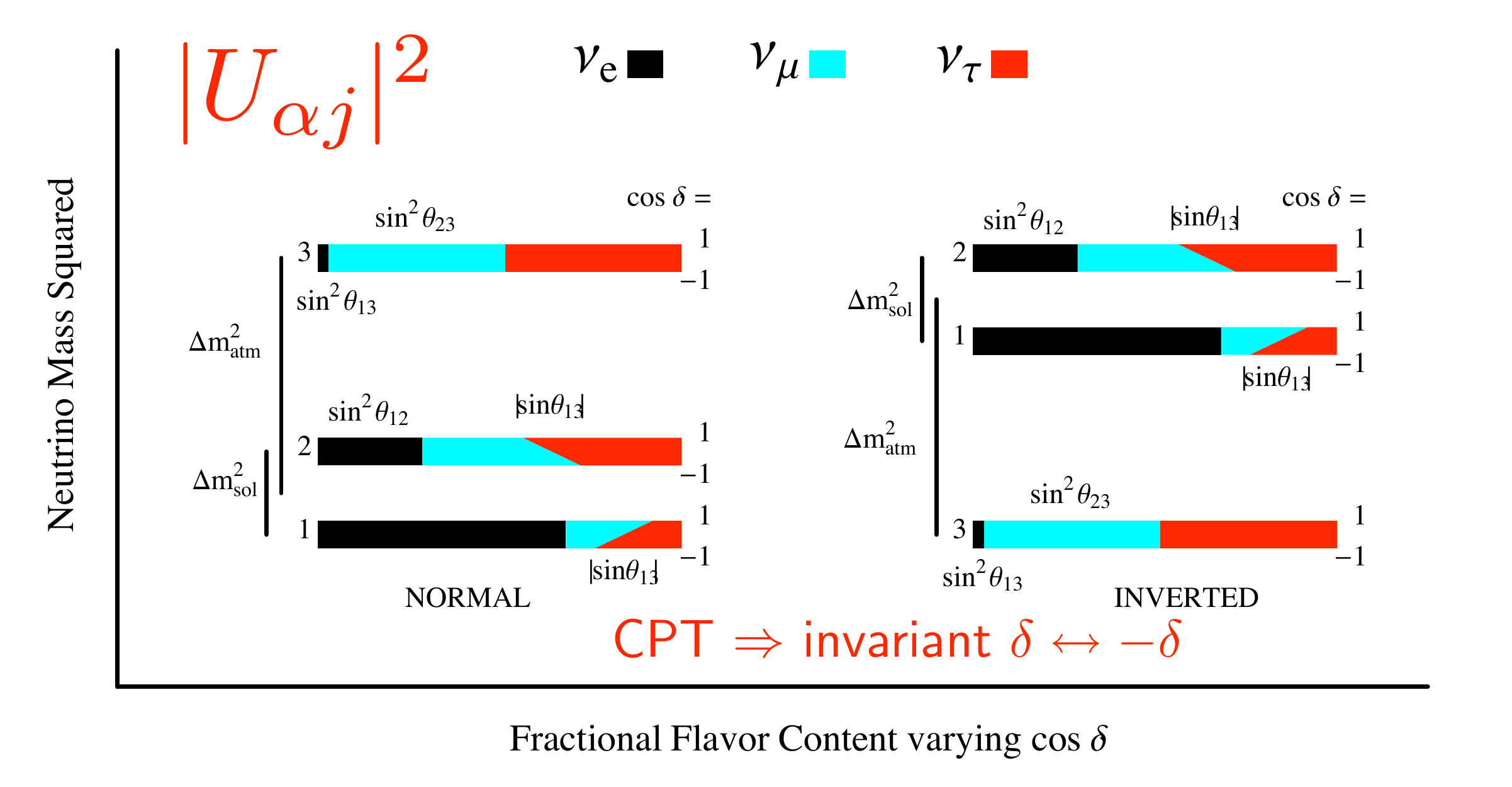}
\end{center}
\caption{The flavor content of the neutrino mass eigenstates\protect\cite{Mena:2003ug}.  The width of the lines is used to show how these fractions change as 
$\cos \delta_{CP}$ varies from -1 to +1.  Of course, this figure must be the same for neutrinos and anti-neutrinos, if CPT is conserved. %In this figure $\Delta m^2_{atm}\equiv \delta m^2_{atm}$ and $\Delta m^2_{sol} \equiv \delta m^2_{sol}$.
}
\label{fig:deltaCP}
\end{figure*}

The standard representation of PMNS mixing matrix is given as follows:
\begin{eqnarray}
U_{e2}=\cos \theta_{13}  \sin \theta_{12}  \nonumber \\
U_{\mu 3}=\cos \theta_{13} \sin \theta_{23} \\
U_{e3} =\sin \theta_{13} e^{-i\delta} \nonumber
\end{eqnarray}
with all other elements following by unitarity. The square of the elements of the PMNS matrix give the fractional flavor content, e.g. $|U_{e2}|^2$ is the fraction of $\nu_2$ that is $\nu_e$.  Fig. \ref{fig:deltaCP} gives this fraction for all the mass eigenstates.

Alternatively, we can write
\begin{eqnarray}
 \hspace*{-2cm} \sin^2 \theta_{13}   \equiv  \vert U_{e3}\vert^ ,  & \quad 
\sin^2 \theta_{12} \equiv \frac{\vert U_{e2}\vert^2}{(1- \vert U_{e3}\vert^2)}  \approx \vert U_{e2}\vert^2,  \quad &
\sin^2 \theta_{23}  \equiv \frac{\vert U_{\mu 3}\vert^2}{(1- \vert U_{e3}\vert^2)} \approx \vert U_{\mu 3}\vert^2 \nonumber  
\end{eqnarray}
 where the $\approx$ follows from  the fact that we know that $ \vert U_{e3}\vert^2\ll1$. 
 
 Our current knowledge of these mixings angles is approximately as follows:
 \begin{eqnarray}
 \sin^2 \theta_{12} \approx \frac{1}{3},   \quad  \sin^2 \theta_{23} \approx \frac{1}{2} , \quad   \sin^2 \theta_{13} \approx 0.02  %\nonumber \\
  \quad {\rm and}  \quad 0 \leq \delta < 2\pi  
 \end{eqnarray}
which are the values used in this figure. For more precise values see the latest PDG.

\subsection{The Neutrino Unitarity Triangle}
\begin{figure*}[t]
\begin{center}
 \includegraphics[angle=0, scale= .63]{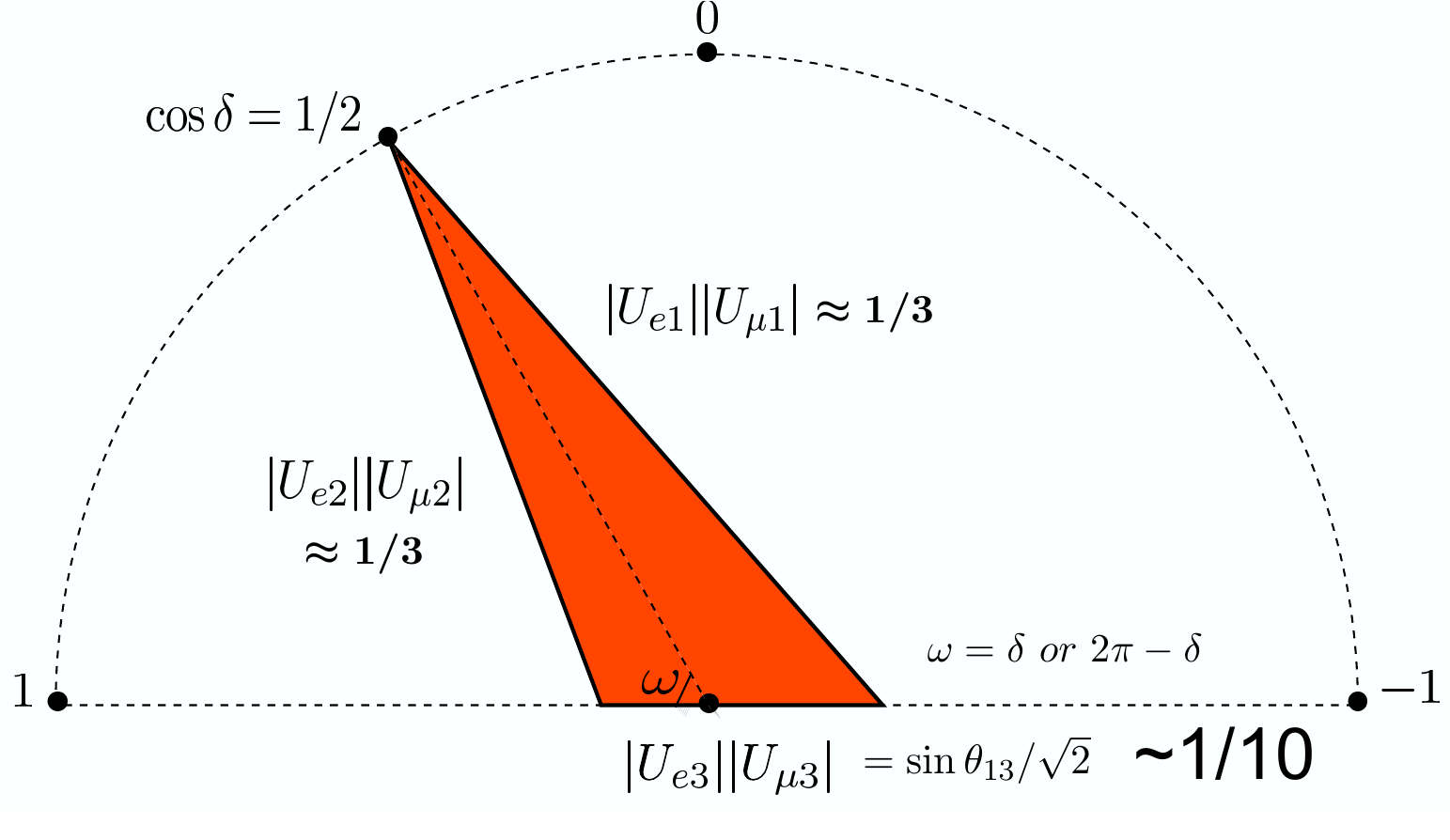}
\end{center}
\caption{The neutrino unitarity triangle \protect\cite{Nunokawa:2007qh} for the first two rows of the PMNS matrix. 
The sides are in the approximate ratio of 3 to 3 to 1 and twice the area of this triangle is the Jarlskog invariant \protect\cite{Jarlskog:1985ht}, which determines the size of CP violation.}
\label{fig:unitarity}
\end{figure*}

The orthogonality of the rows and columns of the PMNS mixing matrix, gives six unitarity relationships, that can be shown as  triangles in the complex plane.  However, only one of these triangles does not involve the $\tau$-neutrino which is experimentally challenging in both detection and production.  This unique unitarity triangle, \cite{Nunokawa:2007qh},  is
given by
 \begin{eqnarray}
U^*_{\mu 1} U_{e 1} +U^*_{\mu 2} U_{e 2}+ U^*_{\mu 3} U_{e 3}=0.
\end{eqnarray}
Where the magnitude of the elements of U are approximately given by
\begin{eqnarray}
\hspace{-2cm} |U_{\mu 1}| \approx \sqrt{\frac{1}{6}}, \quad |U_{e 1}| \approx \sqrt{\frac{2}{3}} 
\quad |U_{\mu 2}| \approx \sqrt{\frac{1}{3}}, \quad |U_{e 2}| \approx \sqrt{\frac{1}{3}} 
\quad |U_{\mu 3}| \approx \sqrt{\frac{1}{2}}, \quad |U_{e 3}| \approx \frac{1}{6}  \nonumber
\end{eqnarray}
and the phases are unknown. 
Thus the size of the sides of this unitarity triangle are 
\begin{eqnarray}
\hspace{-2cm} |U_{\mu1}||U_{e1}| \sim 0.1 - 0.4,  \quad |U_{\mu2}||U_{e2}| \sim 0.2 - 0.4, \quad  
|U_{\mu 3}||U_{e3}| \sim 0.08 - 0.12,
\end{eqnarray}
see Fig.~\ref{fig:unitarity}. To test this relationship one needs to measure $|U_{\mu 1}|$ and $|U_{\mu 2}|$ separately.  Current experiments do not allow this determination.  To separate these two elements one needs, for example, a $\nu_\mu$ disappearance experiment at the solar L/E $\approx$ 15,000 km/GeV !  A $\nu_\mu$ beam to a detector in geosynchronous orbit is one possibility, but at the current time this is science fiction.
Without imposing unitarity, the knowledge of some of the elements of the PMNS matrix is poor.  For example, all of our information on $U_{\tau 1}$ comes entirely from imposing unitarity!

\subsection{Leptons v Quarks}
The Lepton and Quark mixing matrices are very different:
\begin{eqnarray}
\hspace*{-2cm}
U_{PMNS} \sim \left( \begin{array}{ccc}
0.8 & 0.5 & 0.2 \\
0.4 & 0.6 & 0.7\\
0.4 & 0.6 & 0.7 
\end{array} \right)
\quad {\rm and} \quad 
V_{CKM} \sim \left( \begin{array}{ccc}
\approx 1 & 0.2 & 0.001 \\
0.2 & \approx 1 & 0.01\\
0.001& 0.01 & \approx 1 
\end{array} \right).
\end{eqnarray}
The CKM mixing matrix is approximately the identity matrix plus small (Cabibbo) corrections whereas  the Lepton matrix could be some special matrix, bimaximal or tribimaximal, plus small (Cabbibo) corrections. This way of thinking has lead to a number of testable relationships between the mixing angles \cite{Mohapatra:2006gs}, such as:
\begin{eqnarray}
\theta_{13} \approx \theta_c /\sqrt{2} \nonumber \\
\theta_{12} = \theta_{s}+\theta_{13} \cos \delta, \quad {\rm where} ~~\theta_s= 45^\circ, 35^\circ  ~~ {\rm or} ~32^\circ\\
\theta_{23} = 45^\circ+ \kappa \theta_{13} \cos \delta, \quad {\rm where } ~\kappa= \sqrt{2} ~~{\rm or} ~-1/\sqrt{2}. \nonumber
\end{eqnarray}
Although, these models are not completely compelling, the relationships they produce are worth testing as maybe we
will make progress in understanding this exceedingly challenging problem.  Much like the Rutherford-Bohr atom lead to a more complete understanding of the atom with the discovery of quantum mechanics.

\section{Neutrino Phenomenology}

In this section, I will address some important topics in neutrino phenomenology related to disappearance and appearance experiments.

\subsection{Neutrino Disappearance Experiments}
For neutrino disappearance experiments, the vacuum  oscillation probability for $\nu_\alpha =(\nu_e, \nu_\mu, \nu_\tau)$ can be written as \cite{Minakata:2006gq}
\begin{eqnarray}
P(\nu_\alpha \rightarrow \nu_\alpha) & = & 1- 4 |U_{\alpha 1}|^2 |U_{\alpha 2}|^2 \sin^2 \Delta_{21} \\
& & - 4 |U_{\alpha 3}|^2 (1-|U_{\alpha 3}|^2)
 ~ \{ r_a \sin^2 \Delta_{31}+(1-r_\alpha) \sin^2 \Delta_{32} \}  \nonumber \\
 & & \hspace*{-5cm} {\rm where} ~~ \Delta_{ij} =  \frac{\delta m^2_{ij} L}{4E} 
 \quad {\rm and}  \quad {r_\alpha=\frac{ |U_{\alpha 1} |^2}   {(|U_{\alpha 1} |^2 + |U_{\alpha 2} |^2)}}. \nonumber 
\end{eqnarray}
Near the atmospheric first oscillation minimum, $\Delta_{31}\approx \Delta_{32} \approx \pi/2$, this can be approximated by
\begin{eqnarray}
P(\nu_\alpha \rightarrow \nu_\alpha)  & = & 1- \sin^2 2\theta_{\alpha \alpha} \sin^2 \frac{\delta m^2_{\alpha \alpha} L}{4E} +{\mathversion{bold} {\cal O}(\Delta^2_{21})}
\label{eqn:Pdis_eff}  \\[3mm]
& & \hspace*{-4.5cm} {\rm where} %~~\Delta_{\alpha \alpha} =  \frac{\delta m^2_{\alpha \alpha} L}{4E}, 
 \quad \delta m^2_{\alpha \alpha} \equiv r_\alpha |\delta m^2_{31}|+(1- r_\alpha) |\delta m^2_{32}| 
 %\nonumber \\ & & \hspace*{3cm}  
 \quad {\rm and}  \quad
 \sin^2 2\theta_{\alpha \alpha}=4|U_{\alpha 3}|^2(1-|U_{\alpha 3}|^2). \nonumber 
\end{eqnarray}
Any other choice for the effective $\delta m^2$, other than $\delta m^2_{\alpha \alpha}$, induces a
% introduces a 
${\cal O}(\Delta_{21})$ term in eqn.~\ref{eqn:Pdis_eff} and since $\Delta_{21} \approx 1/20$ this reduces the accuracy of the approximation from 0.3\% to 5\%, a significant change.

\begin{figure*}[b]
\begin{center}
 \includegraphics[angle=0, scale= .40]{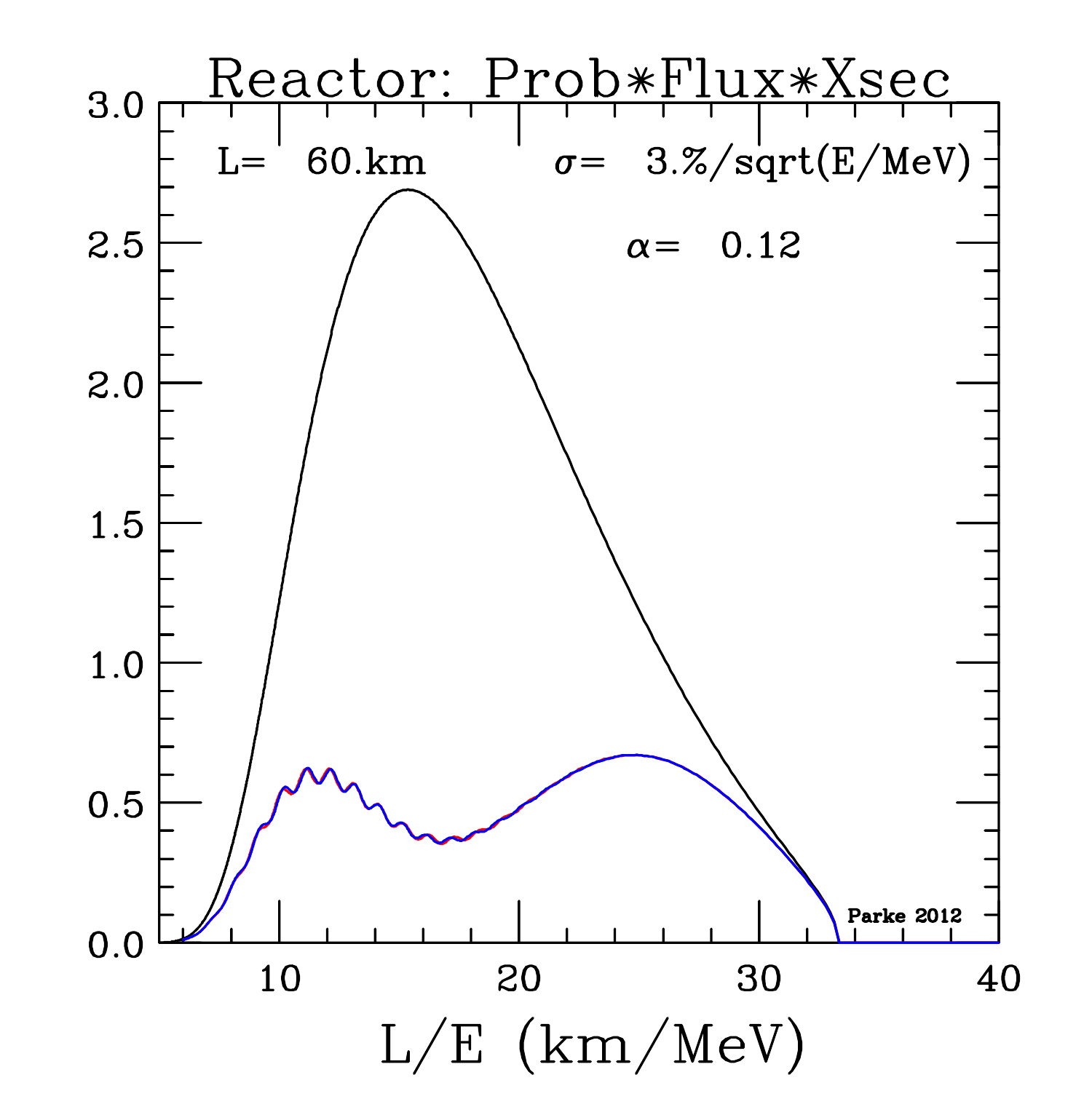}
\end{center}
\caption{The oscillated reactor electron antineutrino flux times cross section for the two mass hierarchies (blue [red] is the normal [inverted] hierarchy)  where the $\delta m^2$'s have been chosen, so as to minimize the difference in the oscillation probabilities between the two hierarchies, but remain within their measurement uncertainties. The black curve is the unoscillated spectrum times cross section and the energy resolution is assumed to be $3\%/\sqrt{E/MeV}$.
}

\label{fig:edisapp}
\end{figure*}

Until your uncertainty on your measurement of P is less than ${\cal O}(\Delta^2_{21}) \sim 0.003$ then
\begin{itemize}
\item three flavor effects are invisible,
\item the effective $\delta m^2$ measured is  $\delta m^2_{\alpha \alpha}= r_\alpha |\delta m^2_{31}|+(1- r_\alpha) |\delta m^2_{32}|$,\\ \indent the $\nu_\alpha$ average of $|\delta m^2_{31}|$ and $|\delta m^2_{32}|$,
\item and the effective $\sin^2 2 \theta$ is $ \sin^2 2\theta_{\alpha \alpha}=4|U_{\alpha 3}|^2(1-|U_{\alpha 3}|^2)$.
\end{itemize}
So the MINOS, T2K and NO$\nu$A $\nu_\mu$ disappearance experiments\footnote{Matter effects are very small in the $\nu_\mu$ disappearance channel at these baselines.} all measure 
\begin{eqnarray}
 & & \delta m^2_{\mu \mu}   =  \frac{|U_{\mu1}|^2 ~|\delta m^2_{31}| + |U_{\mu2}|^2 ~ |\delta m^2_{32}|}
{|U_{\mu1}|^2 + |U_{\mu2}|^2 }  \\
& = & \left\{ (s^2_{12}+s^2_{13} c^2_{12}  t^2_{23} -2 s_{13} t_{23} s_{12} c_{12} \cos \delta)~|\delta m^2_{31}| 
\right.  \nonumber \\
& & \hspace{1cm} \left. + (c^2_{12}+s^2_{13} s^2_{12}  t^2_{23}+2 s_{13} t_{23} s_{12} c_{12} \cos \delta)~|\delta m^2_{32}| \right\}
/(1+s^2_{13} t^2_{23}), \nonumber
\end{eqnarray}
whereas the Daya Bay, RENO and Double CHOOZ experiments measure 
\begin{eqnarray}
\hspace*{-1cm} \delta m^2_{ee} = \frac{|U_{e1}|^2 ~|\delta m^2_{31}| + |U_{e 2}|^2 ~ |\delta m^2_{32}|}
{|U_{e 1}|^2 + |U_{e 2}|^2 } = \cos^2 \theta_{12} ~|\delta m^2_{31}| + \sin^2 \theta_{12} ~ |\delta m^2_{32}|.
\end{eqnarray}
With sub-1\% precision on $\delta m^2_{\mu \mu}$  and  $\delta m^2_{ee}$ the neutrino mass hierarchy can be determined, as
\begin{eqnarray}
 \delta m^2_{ee}  > (<) ~\delta m^2_{\mu \mu}  \quad {\rm Normal ~(Inverted) ~Hierarchy} .
 \end{eqnarray}
  This appears, at this time, to be exceptionally challenging especially determining the absolute energy scale of 
  $\delta m^2_{\mu \mu}$ to this precision.

  Near the first solar oscillation minimum, $\Delta_{21} \approx \pi/2$, the interference between the $\{31\}$ and $\{32\}$
  oscillations leads to an advance (retardation) in the phase of the atmospheric oscillation for the normal (inverted) hierarchy.  The reactor neutrino disappearance experiments could in principle use this to determine the mass hierarchy but this determination is extremely challenging due mainly to not having a  highly precise measurement of either 
$ \delta m^2_{ee} $ or $\delta m^2_{\mu \mu} $.  Without this determination, one has to let the $\delta m^2$'s float between the two hierarchies within the measurement uncertainties and this leads to an obfuscation of the advance or retardation of the phase of the atmospheric oscillations, see Fig. \ref{fig:edisapp}.

Given that it is hard to see the two curves in Fig. \ref{fig:edisapp}, there are important systematic issues, such as the linearity of the detector energy scale, which was first address in \cite{Parke:2008cz} and recently revisited in \cite{Qian:2012xh}
before one can be convinced such a determination of the mass hierarchy can be achieved.  Given the size of the detector 
planned there are ample statistics to make the determination provided that the systematic issues are under control.

\subsection{Neutrino Appearance Experiments}

Genuine three flavor effects, like CP violation, can be observed in  long baseline 
$\nu_\mu \rightarrow \nu_e$ appearance experiments
or in one of its CP or T conjugate channels. That is, in one of following transitions
\def \al{\mu}
\def \be{e}
%For $\alpha \neq \beta$\\[-0.5in]
%\vspace*{1cm}
{\Large
\begin{center}
\hspace*{-0.7cm}\begin{tabular}{lrclr}
& & {CP} & & \\[0.1in]
&$\nu_\al \rightarrow \nu_\be$ &
$\Longleftrightarrow$ &
%$\bar{\nu}_\al \rightarrow \bar{\nu}_\be$  & & ~~~~~~~~~Super-Beams \\[0.2in]
$\bar{\nu}_\al \rightarrow \bar{\nu}_\be$  &  \\[0.2in]
{T} &  $\Updownarrow$ 
%&\quad \quad $\times {\tiny\red{ \rm (CPT)}}$ \quad \quad \quad  & ~~\quad $\Updownarrow$
 \quad \quad
&\quad \quad \quad \quad \quad  & ~~\quad $\Updownarrow$ \quad \quad {T} & \\[0.2in]
&$\nu_\be \rightarrow \nu_\al$ &
$\Longleftrightarrow$ &
%$\bar{\nu}_\be \rightarrow \bar{\nu}_\al$ & & Nu-Factory \\
$\bar{\nu}_\be \rightarrow \bar{\nu}_\al$ &   \\[0.1in]
& & {CP} & & \\[5mm]
\end{tabular}
\end{center}
}
\noindent Processes across the diagonal are related by CPT. The first row will be explored in
very powerful conventional beams, T2K \cite{Itow:2001ee}, NO$\nu$A \cite{Ayres:2002ws}, Superbeams, HyperK \cite{Abe:2011ts}, LBNE \cite{Adams:2013qkq}, ESSnuSB \cite{Baussan:2012cw}, whereas the second row could be explored in Nu-Factories \cite{Apollonio:2012hga} or Beta Beams \cite{Burt:2011zb}.

In vacuum, the probability for $\nu_\mu \rightarrow \nu_e$ is derived as follows, \cite{Cervera:2000kp}, 
\begin{eqnarray}
   \hspace{-0.5cm} P(\nu_\mu \rightarrow \nu_e)   = & &  \hspace{-1.0cm} | ~U_{\mu1}^* e^{-im^2_1L/2E} U_{e1} 
+  U_{\mu2}^* e^{-im^2_2L/2E} U_{e2} 
+  U_{\mu3}^* e^{-im^2_3L/2E} U_{e3}~|^2  \nonumber \\
& = &   | 2U^*_{\mu 3}U_{e3} \sin \Delta_{31} e^{-i\Delta_{32}} + 2 U^*_{\mu2}U_{e2}\sin \Delta_{21}|^2 
\nonumber \\
&\approx & |{\sqrt{P_{atm}}}e^{-i(\Delta_{32}+\delta)} + {\sqrt{P_{sol}}}|^2  \\[5mm]
%\nonumber 
%= P_{atm} + 2\sqrt{P_{atm}} \sqrt{P_{sol}}\cos(\Delta_{32}+\delta) + P_{sol}
\label{eq:Pme}
 \noindent {\rm where} ~~\sqrt{P_{atm}}   & = & 2|U_{\mu 3}||U_{e3}| \sin \Delta_{31}  = \sin \theta_{23} \sin 2 \theta_{13} \sin \Delta_{31} \nonumber \\
{\rm and}  ~~ \sqrt{P_{sol}} & \approx &  \cos \theta_{23} \sin 2 \theta_{12} \sin \Delta_{21}. \nonumber
 \end{eqnarray}
Note, $\sqrt{P_{atm}}$ and $\sqrt{P_{sol}}$ are just  the two flavor oscillation amplitudes at the atmospheric and solar scales, respectively.

\begin{figure}[tb]
\begin{center}
\includegraphics[width=0.35\textwidth]{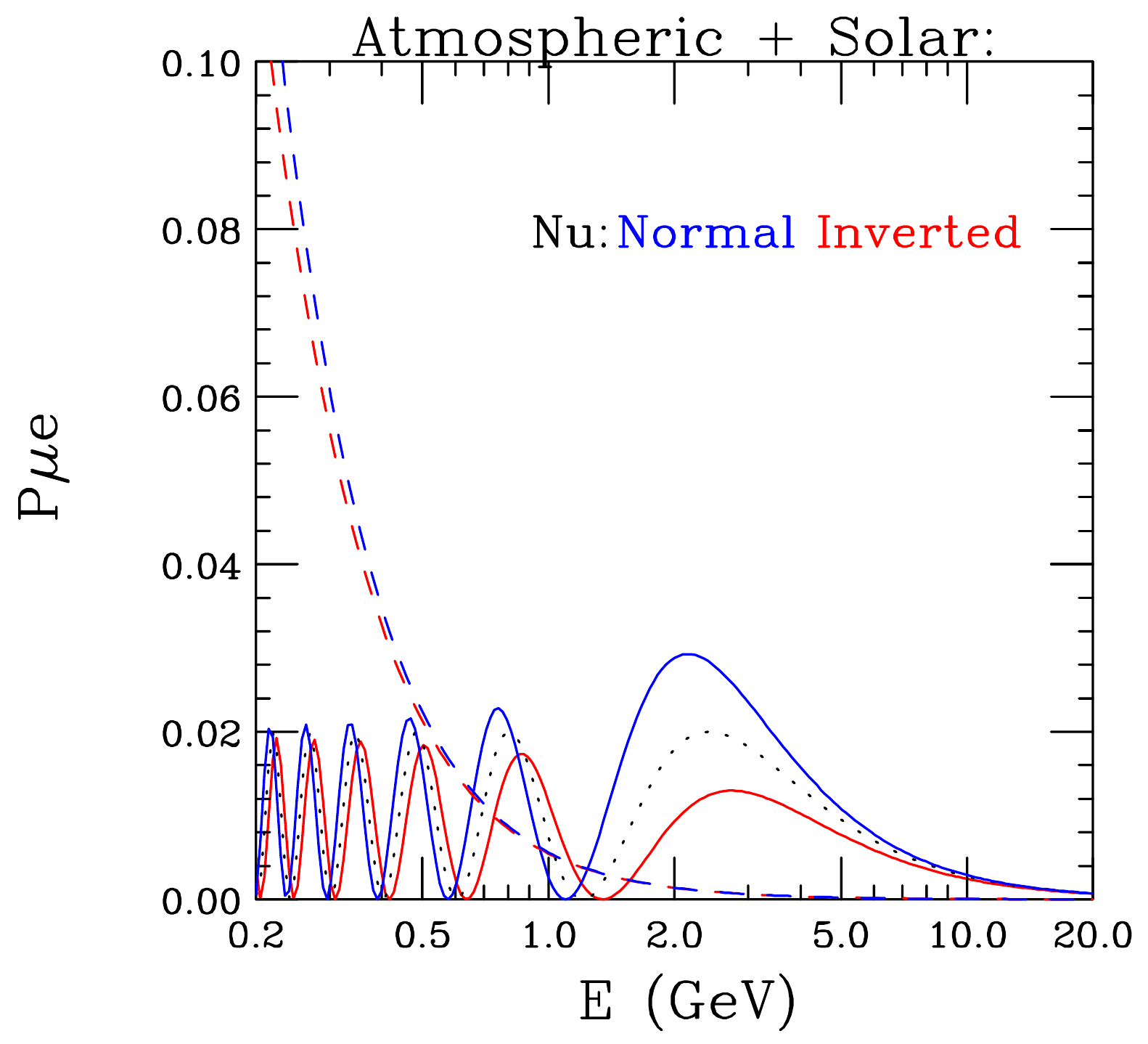}
\hspace*{1.0cm}
\includegraphics[width=0.35\textwidth]{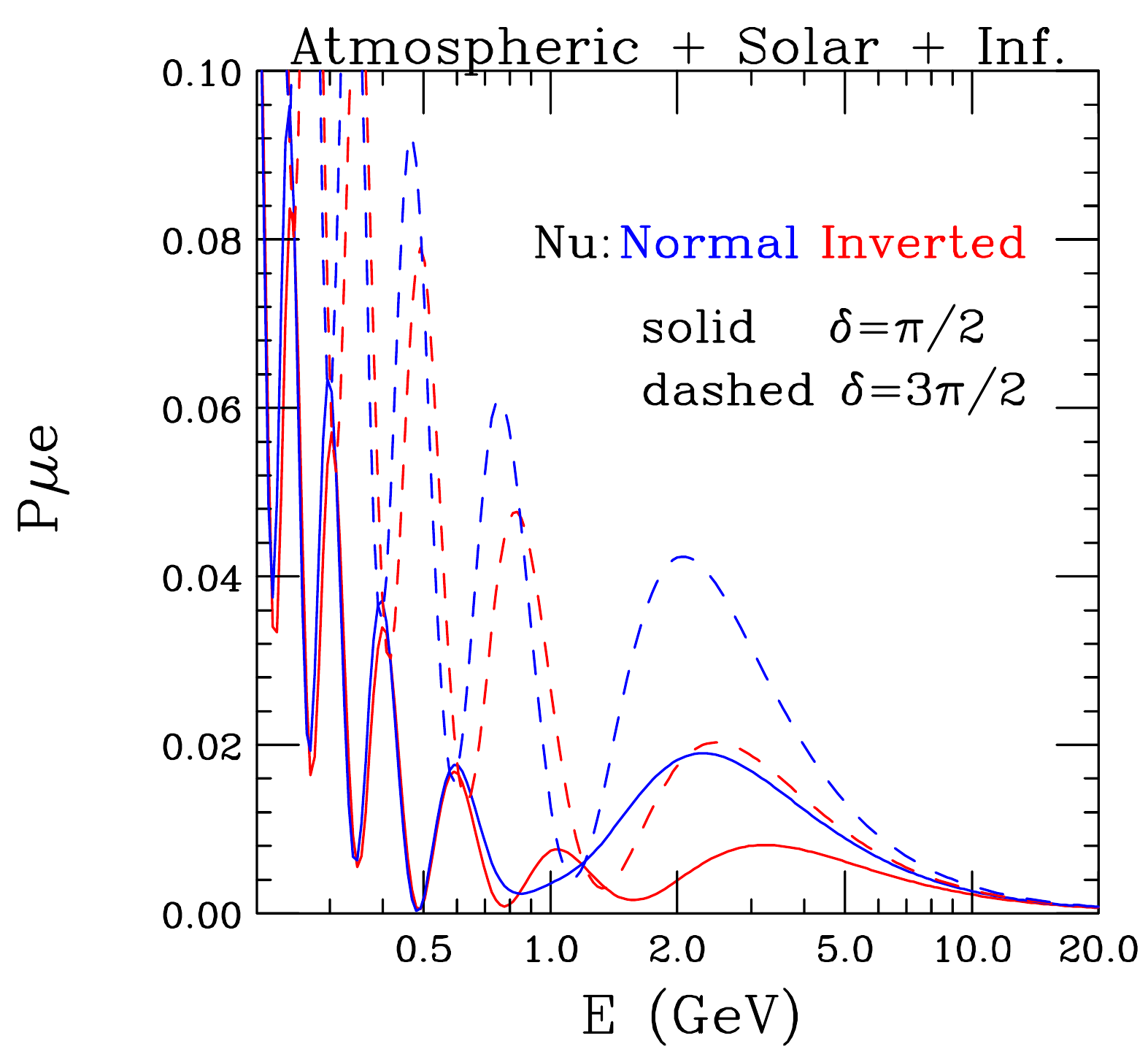}
\end{center}
\caption{The left panel shows the two components $P_{atm}$ and $P_{sol}$ in matter for the normal and inverted hierarchies for $\sin^2 2 \theta_{13} =0.04$ and a baseline of 1200 km. The right panel shows the total probability including the interference term between the two components for various values of the CP phase $\delta$ for the neutrino.  Notice that the coherent sum of two amplitudes shows a rich structure depending on the hierarchy and value of CP phase.
These curves can also be interpreted as anti-neutrino probabilities if one interchanges the hierarchy AND the values of the CP phase.}
\label{fig:pmutoe}
\end{figure}
For anti-neutrinos $\delta$ must be replaced with $-\delta$ and the interference term changes
 \begin{eqnarray}
 2\sqrt{P_{atm}} \sqrt{P_{sol}}\cos(\Delta_{32}+\delta)  & \Rightarrow &  2\sqrt{P_{atm}} \sqrt{P_{sol}}\cos(\Delta_{32}-\delta).  \nonumber
 \end{eqnarray}
 Expanding  $\cos(\Delta_{32}\pm\delta)$, one has a CP conserving part
\begin{eqnarray}
2\sqrt{P_{\rm {atm}}} \sqrt{P_{\rm {sol}}} \cos\Delta_{32}\cos\delta
\end{eqnarray}
 and the CP violating part, where - (+) sign is for the neutrino (anti-neutrino) channel,
\begin{eqnarray}
\mp 2\sqrt{P_{\rm {atm}}} \sqrt{P_{\rm {sol}}} \sin\Delta_{32}\sin\delta \nonumber \\ = \mp \sin \delta \sin 2\theta_{13} \cos \theta_{13} \sin 2 \theta_{12} \sin 2 \theta_{23}  \sin \Delta_{31} \sin \Delta_{32} \sin \Delta_{21} \nonumber \\
= \mp J \sin \Delta_{31} \sin \Delta_{32} \sin \Delta_{21}
\end{eqnarray}
where $J=\sin \delta \sin 2\theta_{13} \cos \theta_{13} \sin 2 \theta_{12} \sin 2 \theta_{23} $ is the Jarlskog invariant \cite{Jarlskog:1985ht}.
 This allows for the possibility  that CP violation maybe able to be observed in the neutrino
 sector, since it allows for $P(\nu_\mu \rightarrow \nu_e) \neq P(\bar{\nu}_\mu \rightarrow \bar{\nu}_e) $ in vacuum.

\begin{figure}[b]
\vspace*{-1cm}
\includegraphics[width=.5\textwidth]{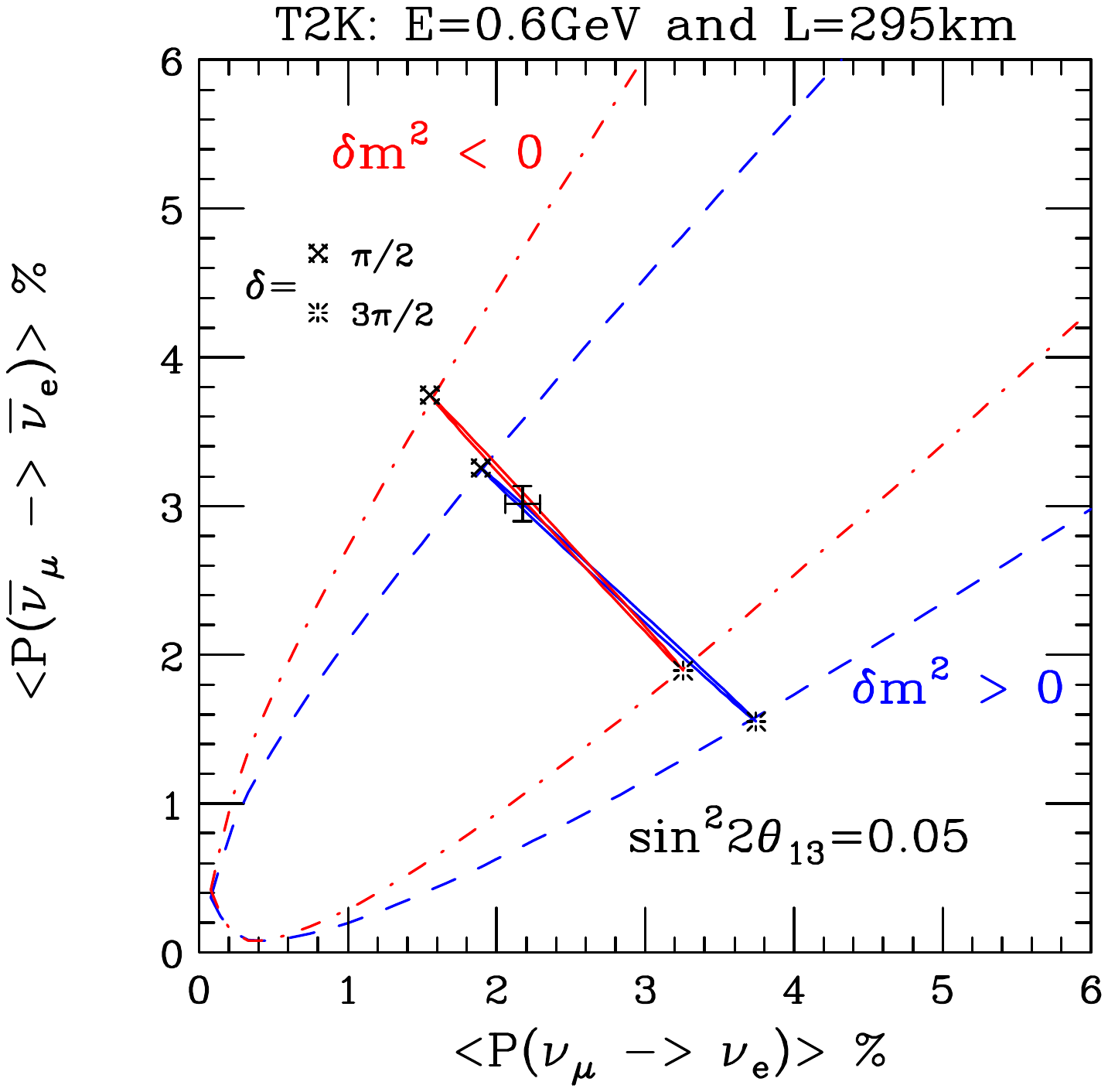}
\includegraphics[width=.5\textwidth]{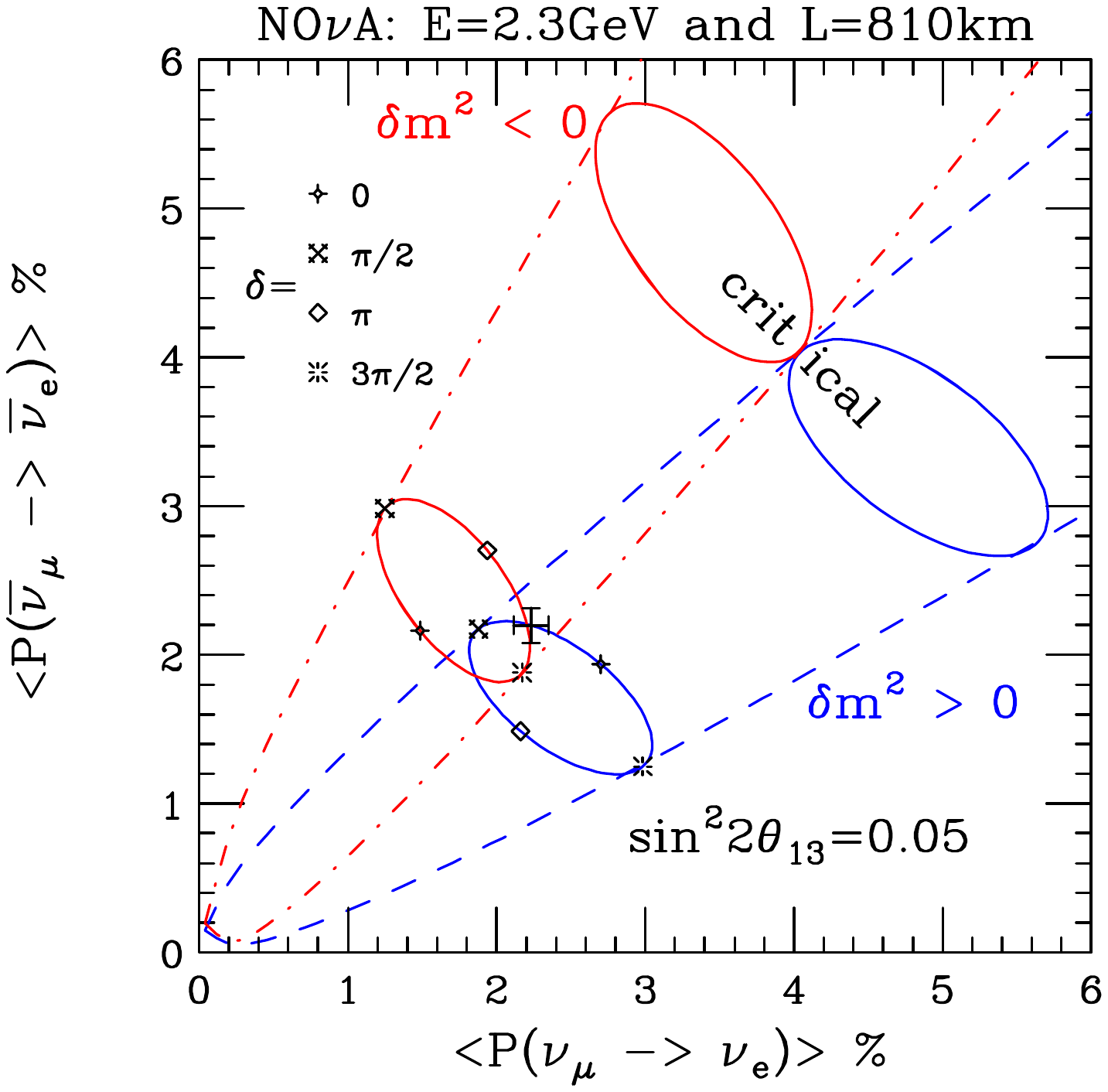}
\vspace*{-3.5cm}
\caption{
The left panel is the bi-probability plot for the T2K/HyperK experiment showing the correlation between neutrino and antineutrino 
$\nu_\mu \rightarrow \nu_e$ probabilities.
The matter effect is small but non-negligible for T2K/HyperK. Whereas the left panel is for the NO$\nu$A experiment where the matter effect is 3 times larger.}
\label{fig:biprob1}
\end{figure}

In matter, the two flavor amplitudes, $\sqrt{P_{atm}}$ and $ \sqrt{P_{sol}}$, are modified as follows
\begin{eqnarray}
\sqrt{P_{atm}}  & \Rightarrow & \sin \theta_{23} \sin 2 \theta_{13} \frac{\sin( \Delta_{31} - aL)}{(\Delta_{31} - aL)}\Delta_{31} \nonumber \\
\sqrt{P_{sol}} &  \Rightarrow & \cos \theta_{23} \sin 2 \theta_{12} \frac{\sin( aL)}{( aL)}\Delta_{21}
\end{eqnarray}
where $a=\pm G_FN_e/\sqrt{2} \approx  \left( \rho Y_e/1.3 ~{\rm g ~cm}^{-3}   \right )(4000~km)^{-1}$ and the sign is positive for neutrinos and negative for anti-neutrinos.
This change follows since in both the (31) and (21) sectors the product $\{\delta m^2 \sin 2 \theta\}$ is 
approximately independent of matter effects.  
Fig.~\ref{fig:pmutoe} shows the $\nu_e$ appearance probability as a function of the energy for a distance of 1200 km.
In Fig.~\ref{fig:biprob1} is the bi-probability
plots for both T2K \cite{Itow:2001ee} (as well as the future possible HyperK \cite{Abe:2011ts}), and NO$\nu$A \cite{Ayres:2002ws}  experiments.  It is possible that these two experiments will
determine the mass ordering, and give a hint of CP violation in the neutrino sector with sufficient statistics.

The critical value of $\tan \theta_{23} \sin \theta_{13}$ at which the bi-probability ellipses for the normal hierarchy and the inverted hierarchy separate is given by \cite{Mena:2004sa}
\begin{eqnarray}
(\tan \theta_{23} \sin 2\theta_{13})^{crit}  & =   & \left\{  \frac{\Delta_{31}^2 \sin 2\theta_{12}}{1-\Delta_{31} \cot \Delta_{31}}  \right\} ~\frac{\delta m^2_{21}}{\delta m^2_{31}} /(aL)  \\ 
& \approx  & ~2.3~\frac{\delta m^2_{21}}{\delta m^2_{31}} /(aL) \quad {\rm at} \quad \Delta_{31}=\pi/2. \nonumber
%
%\hspace*{-2cm}\tan \theta_{23} \sin \theta^{crit}_{13} = \left\{ \frac{\pi^2}{8} ~\frac{\sin 2\theta_{12}}{\tan \theta_{23}}\left[ \frac{4\Delta_{31}^2/\pi^2}{1-\Delta_{31} \cot \Delta_{31}}\right]  \right\} ~\frac{\delta m^2_{21}}{\delta m^2_{31}} /(aL) \approx ~\frac{\delta m^2_{21}}{\delta m^2_{31}} /(aL) \quad {\rm at ~VOM.} \nonumber
\end{eqnarray}
 %at VOM.}
 For the NO$\nu$A experiment, this corresponds to
\begin{eqnarray}
(\tan^2 \theta_{23} \sin^2 2 \theta_{13} )^{crit} = 0.13
\end{eqnarray}
For the measured value of $\sin^2 2 \theta_{13} =0.09$, the ellipse separate when 
%$\tan^2 \theta_{23} >1.37$ or 
$\sin^2 \theta_{23}> 0.58$. 
In the overlap region, the value of $\sin \delta$ for the two hierarchies satisfies the following relationship
\begin{eqnarray}
\langle \sin \delta \rangle _{NH} - \langle \sin \delta \rangle _{IH}  & = & 2( \tan \theta_{23} \sin 2\theta_{13})/(\tan \theta_{23}\sin 2\theta_{13})^{crit} \nonumber \\ 
&\approx &
\left\{\begin{array}{ll} 1.7 \tan \theta_{23} & {\rm NO}\nu{\rm A} \nonumber \\
0.57 \tan \theta_{23} & {\rm T2K/HyperK}.
\end{array} \right.
\end{eqnarray}

It is also worth noting the following, that sum of the neutrino and anti-neutrino probabilities at oscillation maximum
can be directly compared to the value of $\sin^2 2\theta_{13}$ measured by the reactor disappearance experiments:
\begin{eqnarray}
\hspace*{-1.8cm} (P(\bar{\nu}_\mu \rightarrow \bar{\nu}_e) + P(\nu_\mu \rightarrow \nu_e)) |_{\Delta_{31}=\pi/2} = 2 \sin^2 \theta_{23} \sin^2 2\theta_{13} 
+{\cal O}\left(   (aL)\left( \frac{\delta m^2_{21}}{\delta m^2_{31}} \right) \right),
\end{eqnarray}
thus determining the quadrant of $\theta_{23}$.  The difference of these probabilities can be used to determine 
the CP violation phase $\delta$ and the mass hierarchy.

\begin{figure}[h]
\begin{center}
\includegraphics[width=0.45\textwidth]{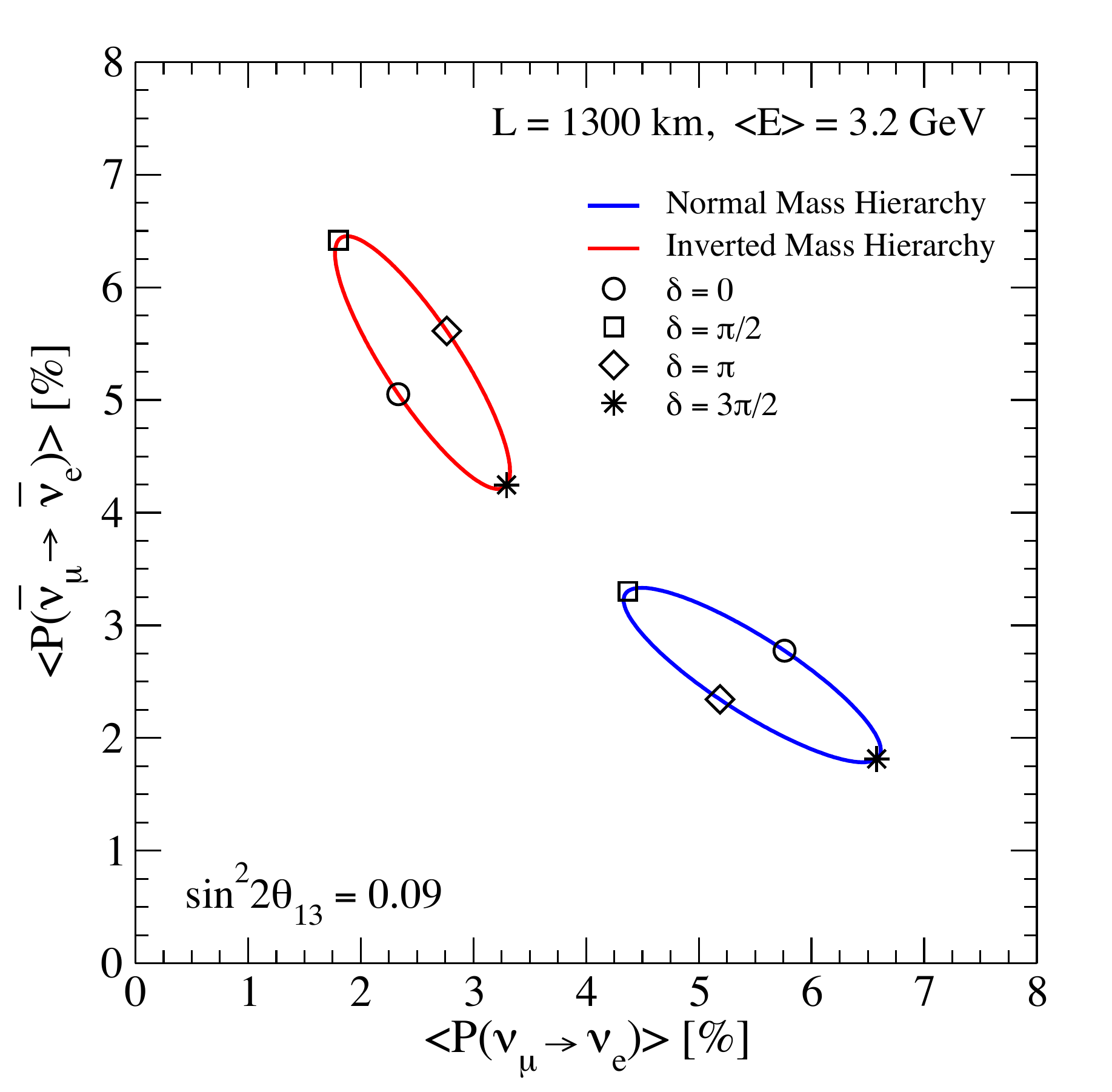}
\end{center}
\caption{The biprobability plot for the LBNE experiment at the same L/E as the NO$\nu$A experiment \protect\cite{Nunokawa}. Notice how widely the normal (blue) and the inverted (red) hierarchies are separated here.
$\sin^2 \theta_{23} = 0.5$ was used for this figure.}
\label{fig:lbne}
\end{figure}

The LBNE experiment \cite{Adams:2013qkq} has a baseline of 1300 km, Fermilab to Homestake, SD which will test the current massive neutrino paradigm in interesting new ways because of its broad band $\nu_\mu$ neutrino beam.  Here the matter effects are larger and the bi-probability ellipses separate at the same L/E as the NO$\nu$A experiment, see Fig.~\ref{fig:lbne}.

\subsection{Asymmetry}
The asymmetry between the neutrino and anti-neutrino appearance probability is defined as \cite{Dick:1999ed} 
\begin{eqnarray}
%\hspace*{-2cm}
A &\equiv &\frac{|P(\nu_\mu \rightarrow \nu_e) -\bar{P}(\bar{\nu}_\mu \rightarrow \bar{\nu}_e)|}
{[P(\nu_\mu \rightarrow \nu_e) +\bar{P}(\bar{\nu}_\mu \rightarrow \bar{\nu}_e)]}, \\
&\approx& \frac{2\sqrt{P_{atm}}\sqrt{P_{sol}}\sin \Delta_{32} \sin \delta}{(P_{atm}+2\sqrt{P_{atm}}\sqrt{P_{sol}}\cos \Delta_{32} \cos \delta+P_{sol})}  \nonumber
\label{eqn:asymmetry}
\end{eqnarray}
In vacuum, the larger this asymmetry the easier it will be to see CP violation.

\begin{figure}[t]
\begin{center}
\includegraphics[width=0.3\textwidth]{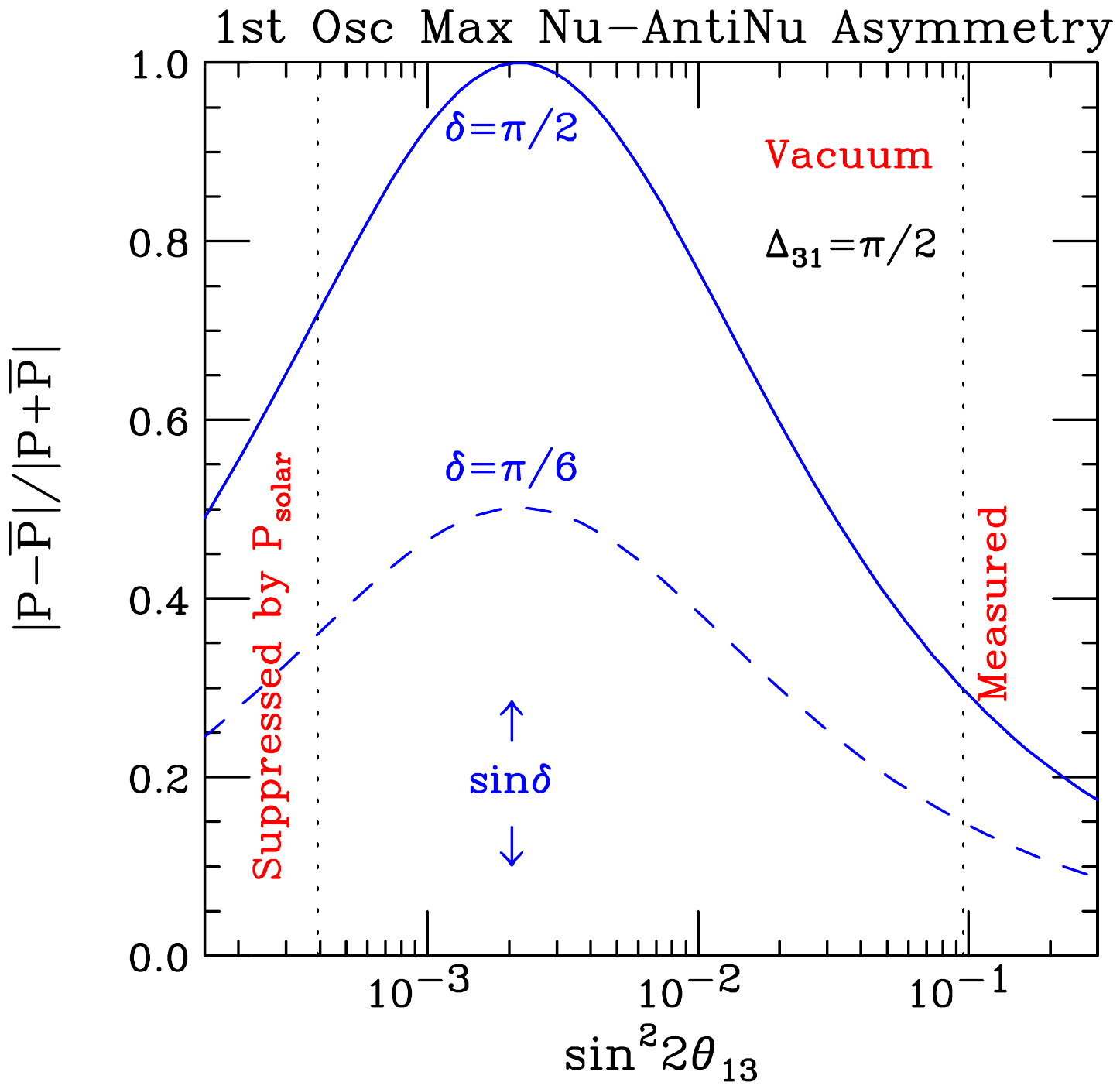}
\hspace*{1.5cm}
\includegraphics[width=0.3\textwidth]{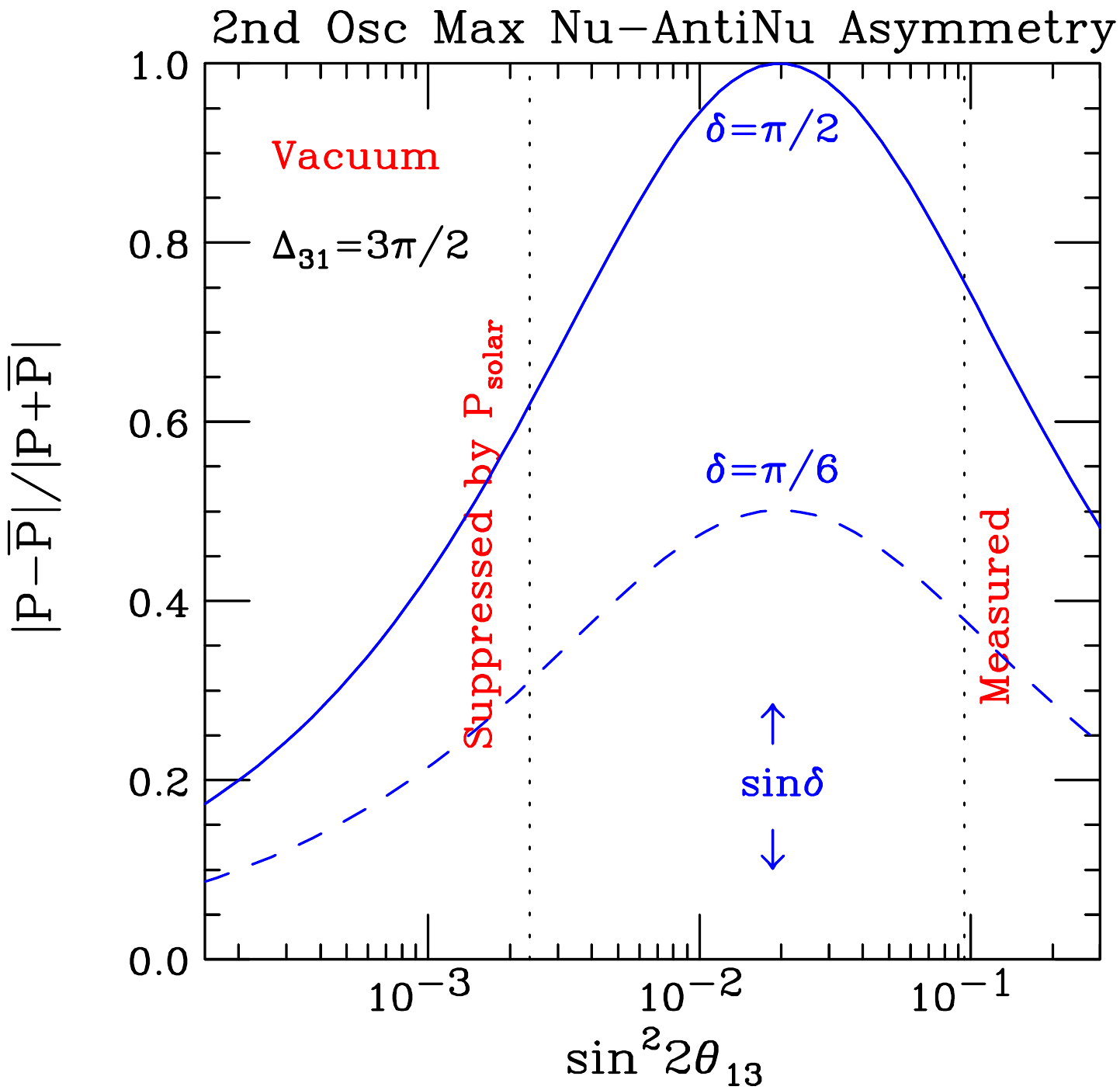}
\end{center}
\caption{The neutrino asymmetry as defined in eqn.~(\protect\ref{eqn:asymmetry}) 
as a function of $\sin^22\theta_{13}$, at the first oscillation maximum \protect\cite{Nunokawa:2007qh} (left panel) and at the second oscillation maximum (right panel) in vacuum. At current measured value of  $\sin^2 2 \theta_{13}=0.090$,  the asymmetries are $A=0.3 \sin \delta$ for the first OM and $A=0.75 \sin \delta$ for the 2nd oscillation maximum.% \protect\cite{Baussan:2012cw}.
}
\label{fig:asymmetry}
\end{figure}

At the first oscillation maximum (OM), as is in the running experiments, T2K and NO$\nu$A and possible future experiments HyperK and LBNE experiments, the vacuum asymmetry is given by
\begin{eqnarray}
A\approx 0.30~ \sin \delta  \quad {\rm at} \quad \Delta_{31}=\pi/2
\end{eqnarray}
which implies that $P(\bar{\nu}_\mu \rightarrow \bar{\nu}_e)$ is between $\frac{1}{2}$ and 2 times $P(\nu_\mu \rightarrow \nu_e)$.  Whereas at the second oscillation maximum, the vacuum asymmetry is
\begin{eqnarray}
A\approx 0.75 ~\sin \delta  \quad {\rm at} \quad \Delta_{31}=3\pi/2
\end{eqnarray}
which implies that $P(\bar{\nu}_\mu \rightarrow \bar{\nu}_e)$ is between $\frac{1}{7}$ and 7 times $P(\nu_\mu \rightarrow \nu_e)$. So that experiments at the second oscillation maximum, like ESSnuSB \cite{Baussan:2012cw}, have a  significantly larger difference between the neutrino and anti-neutrino channels.

\section{The Generalized Intrinsic Degeneracy}
\begin{figure}[h]
\begin{center}
\includegraphics[width=1.03\textwidth]{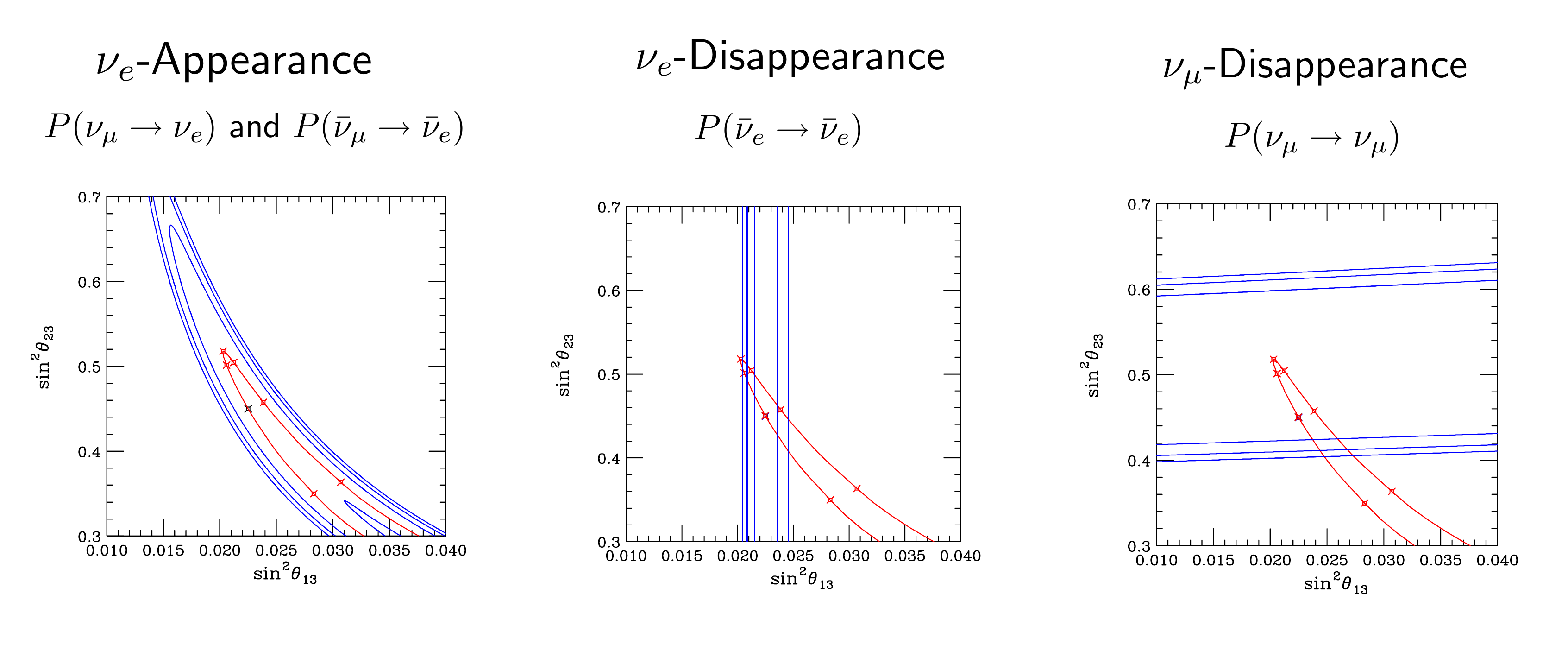}
\end{center}
\caption{The allowed regions in the $\sin^2 \theta_{13} ~v ~\sin^2 \theta_{23}$ plane for the three different types of experiment.  The red line is the exact solution assuming that the input values are  $\sin^2 \theta_{23}=0.45$, $\sin^2 \theta_{13} =0.022$ and $\delta=70^\circ$.  
The marks on this red line indicate values of $\delta$ that are $10^\circ$ apart and the corner is $\delta=90^\circ$.
The blue lines are the 1, 2 ,3 $\sigma$ allowed regions assuming reasonable uncertainties on the measurements.}
\label{fig:intrinsic}
\end{figure}

Let us assume for the moment we known all the parameters governing neutrino oscillation except for  $\sin^2 \theta_{23}$, $\sin^2 \theta_{13}$ and $\delta$ and we will use three different neutrino experiments to determine these parameters \cite{Minakata:2013eoa}:

\begin{itemize}
\item $\nu_\mu \rightarrow \nu_e$ appearance experiments in both the neutrino and antineutrino channels: i.e.
$P(\nu_\mu \rightarrow \nu_e)$ and $P(\bar{\nu}_\mu \rightarrow \bar{\nu}_e)$. In the $\sin^2 \theta_{13} ~v ~\sin^2 \theta_{23}$ plane, these measurements constrain you to a line labelled by the values of $\delta$. See red line in the left panel of Fig. \ref{fig:intrinsic}
\item $\bar{\nu}_e \rightarrow \bar{\nu}_e$ disappearance experiments: $P(\bar{\nu}_e \rightarrow \bar{\nu}_e)$, this measurement determines  $\sin^2 \theta_{13}$ independently of the other variables. Middle panel of Fig. \ref{fig:intrinsic}.
\item $\nu_\mu \rightarrow \nu_\mu$ disappearance experiments: $P(\nu_\mu \rightarrow \nu_\mu)$
this measurement determines the combination of parameters $4\cos^2 \theta_{13} \sin^2 \theta_{23} (1-\cos^2 \theta_{13} \sin^2 \theta_{23})$. Right panel of Fig. \ref{fig:intrinsic}.
\end{itemize}
Also shown in Fig. \ref{fig:intrinsic} is the allowed region for  pseudo-experiments which illustrates the allowed region in the  $\sin^2 \theta_{13} ~v ~\sin^2 \theta_{23}$ for each of these different types of experiment.

Notice the difficult in determining $\sin^2 \theta_{23} \approx 1/2$ and the value of $\delta \approx \pi/2$.  These degeneracies can be solved by information from a sufficient broad neutrino energy spectrum.

\section{Beyond the Neutrino Standard Model}
There are tensions in the $\nu$SM as follows:
\begin{itemize}
\item LSND:  3.8$\sigma$ evidence for anti-$\nu_e$ appearance
\item MiniBooNE: 3.8$\sigma$ combined evidence for $\nu_e$ and anti-$\nu_e$ appearance
\item Reactor: 3.0$\sigma$ evidence for anti-$\nu_e$ disappearance
\item Gallium: 2.7$\sigma$ evidence for $\nu_e$ disappearance
\end{itemize}
This data can be interpreted as the effects of one or more additional sterile neutrinos with a $\delta m^2 \sim 1 ~{\rm eV}^2$. However, there is also tensions with in this extended model between the appearance and disappearance data \cite{Kopp:2013vaa}. There are a number of experiments that are taking data or are planned to address these anomalies. These include
\begin{itemize}
\item Reactor and source experiments looking at the L/E depends for $\bar{\nu}_e \rightarrow \bar{\nu}_e$ and $\nu_e \rightarrow \nu_e$.
\item $\nu_\mu \rightarrow \nu_\mu$ disappearance experiments with both near and far detectors.
\item $\bar{\nu}_\mu \rightarrow \bar{\nu}_e$ or $\nu_e \rightarrow \nu_\mu$ appearance experiments.
\end{itemize}
One of the more ambitious experiments is NuSTORM \cite{Adey:2013afh}, which stores muons in a racetrack shaped ring providing a source of $\nu_e$ with essentially no contamination from other neutrino flavors.  Such an experiment could exclude the LSND allowed region at about 10$\sigma$ and would also be a useful source for measuring neutrino cross sections as the neutrino flux can be calculated from the decaying muon flux with high precision. Such an experiment is a first step on the way to a Neutrino Factory \cite{Apollonio:2012hga} and maybe a Muon Collider \cite{Alexahin:2013ojp} in the future.

\section{Conclusions}
If neutrinos are Majorana in nature and CP violation is observed in neutrino oscillation then the credibility of Leptogenesis will be  greatly enhanced. 
Neutrino oscillation experiments can not only measure CP violation but can also determine whether the 
atmospheric mass hierarchy is normal or inverted and can determine whether the $\nu_\mu$ flavor content is more or less than the $\nu_\tau$ content for the neutrino mass eignestate with the smallest amount of $\nu_e$. The precise measurement of the neutrino mixing and mass parameters will allow us to test the various models predicting these parameters and may lead to a more complete understanding of this notoriously difficult physics problem. If the mass of the lightest neutrino, is significantly smaller than the square root of the solar $\delta m^2$ then there is a new scale in particle physics that needs to be explained. Finally, neutrinos have surprised us in the past and are expected to do so in the future.  Where are these surprises?  Are there light sterile neutrinos?  Do neutrinos decay? What is the size of non-standard interactions? Will LHC find new physics related to neutrino mass? Only the results from further experiments will provide us the answer to these most important questions!

\section{Acknowledgments} 
I wish to thank the Organizers and especially Prof.~Tord~Ekel\"{o}f for this wonderful Symposium and the Nobel Symposia Fund for making this symposium possible.
I also thank all of my collaborators in neutrino physics who through our discussion and paper writing have helped me
better understand the nature of the neutrino.
The author acknowledges partial support from the European Union FP7 ITN INVISIBLES (Marie Curie Actions, PITN- GA-2011- 289442).
Fermilab is operated by the Fermi Research Alliance under contract no. DE-AC02-07CH11359 with the U.S. Department of Energy.

\end{document}